\documentclass[a4paper]{jpconf}

\usepackage{amsmath}
\usepackage{graphicx}

\begin{document}
\address{\small Talk presented at the APS Division of Particles
and Fields Meeting (DPF 2017)\\ July 31-August 4, 2017, Fermilab. C170731}

\title{Automated proton track identification in MicroBooNE using gradient boosted decision trees}

\author{Katherine Woodruff}

\address{on behalf of the MicroBooNE collaboration}
\address{\textit{\\ Physics Department, New Mexico State University \\ Las Cruces, NM 88003}}
\ead{kwoodruf@nmsu.edu}
\begin{abstract}
MicroBooNE is a liquid argon time projection chamber (LArTPC) neutrino
experiment that is currently running in the Booster Neutrino Beam at Fermilab.
LArTPC technology allows for high-resolution, three-dimensional representations
of neutrino interactions. A wide variety of software tools for automated
reconstruction and selection of particle tracks in LArTPCs are actively being
developed. Short, isolated proton tracks, the signal for low-momentum-transfer
neutral current (NC) elastic events, are easily hidden in a large cosmic
background. Detecting these low-energy tracks will allow us to probe
interesting regions of the proton's spin structure. An effective method for
selecting NC elastic events is to combine a highly efficient track
reconstruction algorithm to find all candidate tracks with highly accurate
particle identification using a machine learning algorithm. We present our work
on particle track classification using gradient tree boosting software
(XGBoost) and the performance on simulated neutrino data.
\end{abstract}

\section{Introduction \label{intro}}
The three up and down valence quarks in the nucleon only account for a small
percent of its mass. Gluons that bind the valence quarks split into
quark-antiquark pairs of up, down, and strange flavor. This sea of quarks and
gluons carries the remainder of the nucleon mass. The structure of the
quark-gluon sea and how its elements combine with the valence quarks to give
the nucleon its measured structure is not precisely known.

The net spin of the proton comes from a combination of the spin and orbital
momentum of the quarks and gluons. The net contribution from the spin of
strange quarks and antiquarks, $\Delta s$, is defined as
\begin{equation*}
  \Delta s = \int_0^1 \Delta s(x) \, dx
\end{equation*}
\begin{equation*}
  \Delta s(x) = \sum_{r=\pm 1} r[s^{(r)}(x) + \bar{s}^{(r)}(x)] \,,
\end{equation*}
where $s(\bar{s})$ is the spin-dependent parton distribution function of the
strange (anti)quark, $r$ is the helicity of the quark relative to the proton
helicity and $x$ is the Bjorken scaling variable~\cite{Alberico01}.  In the
static quark model this value is zero.

In the 1980s the European Muon Collaboration~\cite{Ashman89} and several
subsequent experiments found that the Ellis-Jaffe Sum Rule was violated in
polarized, charged-lepton, inclusive, deep inelastic scattering (DIS). The
Ellis-Jaffe sum rule~\cite{Ellis74} assumes that SU(3) flavor symmetry is valid and that
$\Delta s = 0$. For the results to be consistent with exact SU(3) flavor
symmetry, $\Delta s$ must be \textit{negative}. Follow-up measurements using
semi-inclusive deep inelastic scattering have been consistent with $\Delta s =
0$, but these determinations of $\Delta s$ are highly dependent on the
fragmentation functions used~\cite{Aidala12}.

An independent determination of $\Delta s$ can be made using neutral-current
(NC) elastic neutrino-proton scattering. The NC elastic cross section depends
directly on $\Delta s$ and no assumptions about SU(3) flavor symmetry or
fragmentation functions are needed.

Previous measurements using elastic neutron-proton
scattering~\cite{Ahrens86,Garvey93,Aguilar10} have been able to resolve final
state protons down to a kinetic energy of $T\sim~240$~MeV which corresponds to
a momentum transfer of $Q^2=0.45$~GeV$^2$. These measurements also found
$\Delta s < 0$, but the results are highly dependent on the choice of the axial
form factor $Q^2$ dependence. To extract $\Delta s$, $G_A^s$ must be
extrapolated to $Q^2 = 0$. Detecting events with lower momentum transfer would
lessen the dependence on the choice of the model.

Global fits to electron-proton and neutrino-proton elastic scattering data have
found $\Delta s = -0.30 \pm 0.42$~\cite{Pate13}. Based on data from a simulation
of the MicroBooNE detector and the BNB beam, the uncertainty on the global fit
to $\Delta s$ is estimated to decrease by a factor of ten when including
MicroBooNE data.

\section{Elastic neutrino-proton scattering \label{elnup}}
The elastic lepton-nucleon scattering cross section depends on the axial,
electric, and magnetic form factors which represent the finite structure of the
nucleon. The axial form factor, $G_A$, represents the spin structure, and the
electric and magnetic form factors, $G_E$ and $G_M$, represent the electric and
magnetic structure, respectively.

\subsection{Neutral-current elastic scattering \label{ncel}}
The NC elastic neutrino-proton cross section~\cite{Alberico01} can be written as
\begin{equation*}
  \begin{split}
    \left(\frac{d\sigma}{dQ^2}\right)_{\nu}^{NC} &=\frac{G_F^2}{2\pi} \left[
    \frac{1}{2}y^2(G_M^{NC})^2 \right. \\
    &\left. +\left(1-y-\frac{M}{2E}y \right)
    \frac{(G_E^{NC})^2+\frac{E}{2M}y(G_M^{NC})^2}{1+\frac{E}{2M}y}
    \right.\\
    &+ \left. \left(\frac{1}{2}y^2 + 1 - y + \frac{M}{2E}y
    \right)(G_A^{NC})^2 \right. \\
    &+ \left. 2y \left(1-\frac{1}{2}y \right)
    G_M^{NC}G_A^{NC} \right] \,,
  \end{split}
\end{equation*}
where $G_F$ is the Fermi constant, $M$ is the mass of the nucleon, $E$ is the
neutrino energy, $Q^2$ is the momentum transfer, and $y$ is the fractional
energy loss of the incoming lepton.

The neutral-current form factors, $G_A^{NC}$, $G_E^{NC}$, and $G_M^{NC}$, are
functions of $Q^2$ and can all be written as a linear combination of the
individual quark contributions
\begin{equation*}
  \begin{split}
    G_{E,M}^{NC,p}(Q^2) &= \left(1-\frac{8}{3}\textrm{sin}^2\theta_W\right)G_{E,M}^u(Q^2) \\
    &+ \left(-1+\frac{4}{3}\textrm{sin}^2\theta_W\right)G_{E,M}^d(Q^2) \\
    &+ \left(-1 + \frac{4}{3}\textrm{sin}^2\theta_W\right)G_{E,M}^s(Q^2) \\
    G_A^{NC,p}(Q^2) &= \frac{1}{2}\left[-G_A^u(Q^2) + G_A^d(Q^2) + G_A^s(Q^2) \right] \,.
  \end{split}
\end{equation*}
The up, down, and strange quark contributions to the electric and magnetic form
factors of the proton have been determined in a world-wide measurement program
of elastic electron-proton scattering using hydrogen targets and quasi-elastic
electron-nucleon scattering using light nuclear targets (specifically deuterium
and helium)~\cite{Armstrong12,Cates11}.

We plan to measure the ratio of the neutral-current elastic cross section to
the charged-current elastic cross section. The charged-current (CC) elastic
cross section does not depend on $\Delta s$, but it is better known than the NC
elastic cross section. Taking the ratio of the two cross sections reduces
systematic uncertainty on our measurement due to the beam flux, detector
efficiency, and nuclear effects and final state interactions in argon nuclei.

\subsection{Axial form factor \label{axial}}
At the limit when the momentum transfer ($Q^2$) goes to zero, the quark
contributions to the axial form factor become the net contribution of
individual quark spin to the proton spin,
\begin{equation*}
  G_A^q(Q^2 = 0) = \Delta q \qquad (q=u,d,s) \,,
\end{equation*}
so that
\begin{equation*}
  G_A^{NC}(Q^2 = 0) = \frac{1}{2}(-\Delta u + \Delta d + \Delta s) \,.
\end{equation*}
The difference of the up and down spin contributions, $\Delta u - \Delta d$, is
proportional to the axial vector coupling constant $g_A$ measured in hyperon
$\beta$ decay~\cite{Olive16}, therefore a measurement of $G_A^{NC}$ can
determine $\Delta s$.

\subsection{Experimental measurement}

The final state of an NC elastic neutrino-proton interaction consists of a
neutrino and a proton. Since it isn't possible to detect the outgoing neutrino,
the signal is a single proton track. In order to extrapolate the axial form
factor to zero, we need to detect very low energy protons. The kinematics of
the interaction are determined entirely by the proton kinetic energy, $T_P$,
\begin{equation*}
  Q^2 = 2T_P M \,.
\end{equation*}

We estimate that MicroBooNE can detect NC elastic events down to a minimum of
$Q^2 \sim 0.08$ GeV$^2$.  The momentum transfer is determined by the kinetic
energy of the proton in NC elastic interactions. MicroBooNE can detect protons
with a track length of at least 1.5 cm which corresponds to a kinetic energy of
$\sim$40 MeV in liquid argon giving $Q^2 \sim 0.08$~GeV$^2$.

\section{MicroBooNE \label{uboone}}
\begin{figure}
  \includegraphics[scale=0.15]{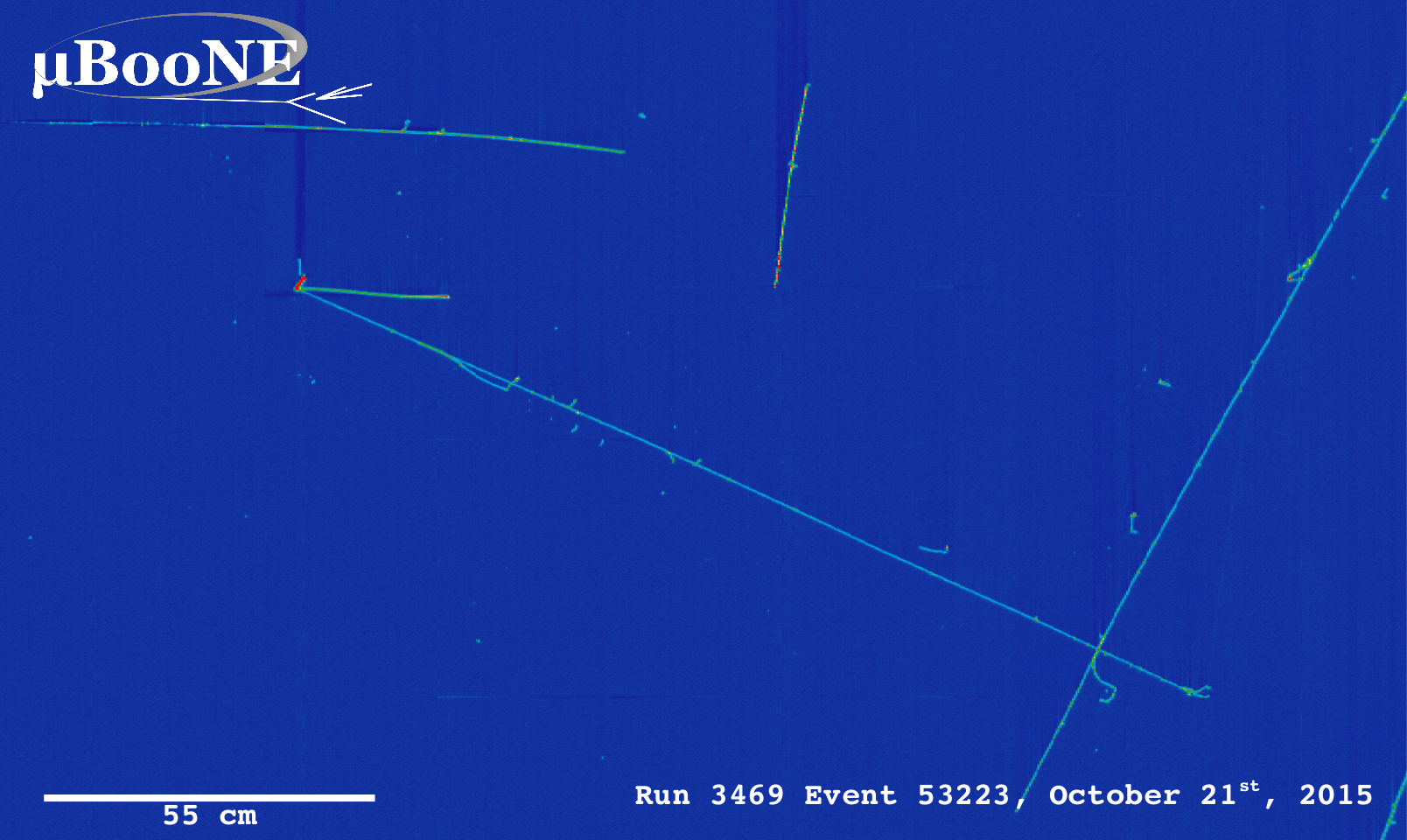}\hspace{2pc}%
  \begin{minipage}[b]{14pc}%
  \caption{A neutrino interaction in the MicroBooNE detector. This is a
  candidate charged-current, muon-neutrino event with a long muon track, a
  charged pion track, and a short proton track coming from the interaction
  vertex.
  \label{fig:event}}
  \end{minipage}
\end{figure}

The MicroBooNE detector~\cite{UBdetector} is a liquid-argon time projection
chamber (TPC) located in the Booster Neutrino Beam at Fermilab. MicroBooNE is a
high-resolution detector designed to be able to accurately identify low-energy
neutrino interactions. It began taking data in October of 2015. Figure~\ref{fig:event}
shows an example neutrino interaction in MicroBooNE.

The MicroBooNE TPC~\cite{UBdetector} has an active mass of 89 tons of liquid
argon. It is 10 meters long in the beam direction, 2.3 meters tall, and 2.5
meters in the electron drift direction. It takes 2.3 ms for electrons to drift
across the full width of the TPC at the operating electric field of 273 V/cm.
Events are read out on three anode wire planes with 3 mm spacing. In addition
to the TPC, there is a light collection system which consists of 32 8-inch PMTs
with nanosecond timing resolution.  The PMTs determine the initial time of the
interaction to help with cosmic rejection. In order for an event to be read
out, there must be an optical signal within a 23 $\mu$s window around the BNB
spill.

The data from each neutrino event in MicroBooNE can be visualized as a set of
three high resolution images (one from each anode plane). Each image has
approximately 20 million pixels (3,000 wires by 9,600 time ticks). It takes
30~MB of disk space to store one MicroBooNE event.

\subsection{Events in MicroBooNE}
An \textit{event} in MicroBooNE has 4.8 ms of TPC readout information. This
includes the 2.3 ms of time after the optical trigger to allow electrons to
drift the entire distance and time before and after the 2.3 ms time frame to
help identify cosmic background. In Fig.~\ref{fig:event}, the X-axis
corresponds to wire number and the Y-axiscorresponds to time tick. The dark
blue background corresponds to no signal on the wire, and the colored pixels
correspond to charge deposited on the wire. The color or intensity of the pixel
corresponds to the amount of energy deposited in the TPC.

\section{Automated event selection \label{evtsel}}
\begin{figure}
  \includegraphics[scale=0.285]{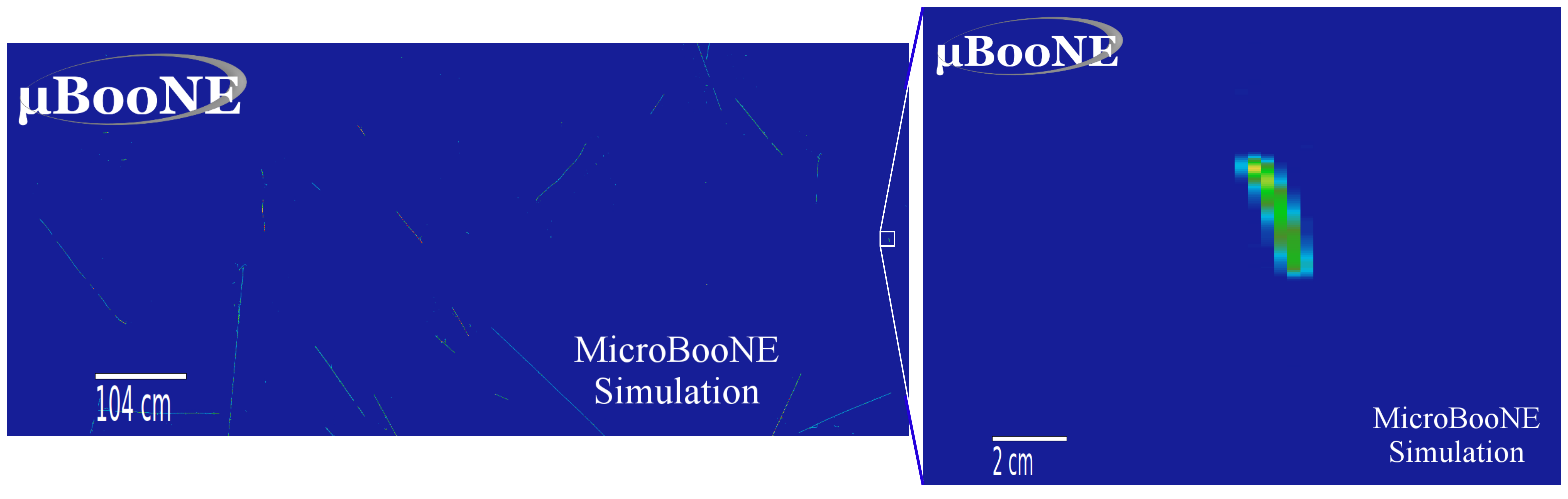}
  \caption{2D event display of a simulated neutral-current elastic event in
  MicroBooNE that was successfully classified as a proton. The top image is a
  close-up event display of the simulated proton track. The bottom image shows
  the side view of the entire MicroBooNE TPC. All of the additional tracks are
  from cosmic rays.  \label{fig:mcevd}}
\end{figure}

MicroBooNE is close to the surface of the Earth, which results in a large
cosmic ray background. Each triggered event is read out for 4.8 ms
(approximately twice the electron drift time), and there are an average of
twelve cosmic muon tracks per readout frame~\cite{UBcosmic}.  This can be seen
in the bottom image in Fig.~\ref{fig:mcevd}. In addition, there are
approximately five times as many event triggers caused by cosmic rays
coincident with the BNB spill than actual neutrino interactions. During
MicroBooNE's three year run, we expect to have $\sim$200,000 neutrino
interactions and $\sim$1,000,000 cosmic interactions.  This means that
automated neutrino event reconstruction and identification algorithms are
required. These algorithms are currently being developed for liquid argon TPCs.

Because each event in MicroBooNE contains around 60 million pixels, we need to
reduce the amount of data without removing a lot of information before trying
to classify the event. We do this by grouping the one-dimensional hits on the
wires into two-dimensional lines on the planes and then into three-dimensional
track objects in the TPC.

\subsection{Track reconstruction in LArSoft \label{larreco}}

Track reconstruction is handled in the Liquid Argon Software framework
(LArSoft)~\cite{Church13}. The three main stages of reconstruction in LArSoft are hit finding,
track finding, and event identification.

One-dimensional hits are found by fitting Gaussian functions to noise-filtered~\cite{UBnoise}
waveforms that are read out from the anode wires in the TPC. This is done for
all of the wires on all three of the planes. The result is a two-dimensional
image for each of the three wire planes, where the two dimensions are wire
number and time. These 2D hits are used as inputs to the Pandora Software
Development Kit~\cite{UBpandora}.  Pandora contains pattern recognition algorithms that have
been optimized to reconstruct tracks from neutrino interactions in liquid argon
TPCs at the BNB energy range.  The Pandora algorithms take a set of hits and
reconstruct neutrino interaction vertices.

At this point the size of each event has been reduced from millions of pixels
to about 20 reconstructed track objects with very little information loss. We
can attempt to identify the type of particle and interaction that produced the
set of tracks. In the NC elastic case, we want to specifically select proton
tracks.

\subsection{Proton track identification \label{protonid}}

Neutral-current elastic interactions are the most difficult to detect
automatically because there is only one visible particle coming from the
interaction vertex. There is no unique topology separating these events from
the cosmic background.

Each reconstructed track object in a MicroBooNE event has several reconstructed
physics properties associated with it. These properties fall into the
categories of geometric, calorimetric, and optical. The geometric properties
are related to the position, shape and size of the track. This includes
variables like whether the track is entering the TPC, how long the track is,
and how curvy a track is. Calorimetric properties all have to do with the
charge deposited along the track. We can use information about how much total
charge was deposited by the track, the average charge deposited per cm, and the
difference between the amont of charge deposited at the beginning or the end of
the track. We can also create variables that represent the shape and scale of
the dE/dx (or dQ/dx) curve at any point along the reconstructed track.
Additionally, we can use optical information from the PMT system to help
characterize tracks. In this work we use the distance between the reconstructed
track and the closest optical flash that was in the beam time window.

\subsubsection{Gradient decision tree boosting}
To identify proton tracks, we use a gradient-boosted decision tree classifier.
We chose to use decision trees because they are easily interpretable and the
inputs can be a mix of numeric and categorical variables. Below is a short
description of gradient tree boosting. A more detailed description can be found
in the documentation for the XGBoost\cite{Chen16} software library that was
used. 

A decision tree can be thought of as a series of if/else statements that
separate a data set into two or more classes. The goal of each cut is to
increase the information gain. For numerical variables any cut value can be
selected by the tree.  At each node of the tree, a split is chosen to maximize
information gain until a set level of separation is reached.  At the terminus
of the series of splits, called a leaf, a class is assigned.

Two weaknesses of decision trees are their tendency to over fit the training
data and the fact that the output is a class label and not a probability.
Gradient-boosting addresses both of these issues by combining many weak
classifiers into a strong one. Each weak classifier is built based on the error
of the previous one. For a given training set, whenever a sample is classified
incorrectly by a tree, that sample is given a higher importance when the next
tree is being created.  Mathematically, each tree is training on the gradient
of the loss function. After all of the trees have been created, each tree is
given a weight based on its ability to classify the training set, and the
output of the gradient-boosted decision tree classifier is the probability that
a sample is in a given class.

\subsubsection{The decision tree model}
We created a multi-class gradient-boosted decision tree classifier, using the
XGBoost software library, to separate five different track types: any proton
track, muons or pions from BNB neutrino interactions, tracks from
electromagnetic showers from BNB interactions, and any non-proton track
produced by a cosmic ray interaction. The classifier takes reconstructed track
features as input and outputs a probability of the track having been produced
by each of the given particle types. The reconstructed features are based on
the track's geometric, calorimetric, and optical properties.

The training data that we use to make the decision trees comes from Monte Carlo
simulation. The BNB interactions are simulated using the GENIE neutrino
generator~\cite{Andreopoulos09}, and cosmic interactions are simulated using
the CORSIKA cosmic ray generator~\cite{Heck98}. The particles generated by
GENIE and CORSIKA are passed to Geant4~\cite{Agostinelli02} where they are
propagated through a simulated MicroBooNE detector. For training and testing of
the trees we only use tracks that were reconstructed in LArSoft.

\begin{figure}
  \centering
  \includegraphics[scale=1.]{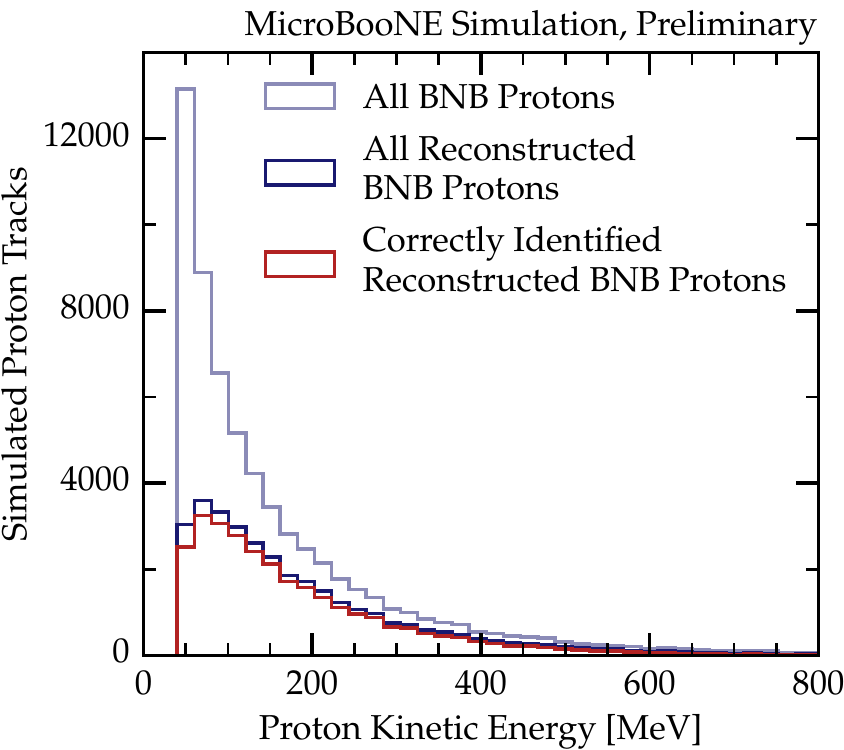}\hspace{2pc}%
  \begin{minipage}[b]{14pc}%
  \caption{Number of simulated proton tracks as a function of true simulated
  kinetic energy is shown. The light blue line shows the total number of
  protons from simulated BNB neutrino interactions. The dark blue line shows
  the total number of those tracks that were reconstructed with the Pandora
  algorithms. The red line shows the subset of the reconstructed tracks that
  are classified as protons by the boosted decision trees. \label{fig:effvke}}
  \end{minipage}
\end{figure}

\begin{figure}
  \centering
  \includegraphics[scale=1.]{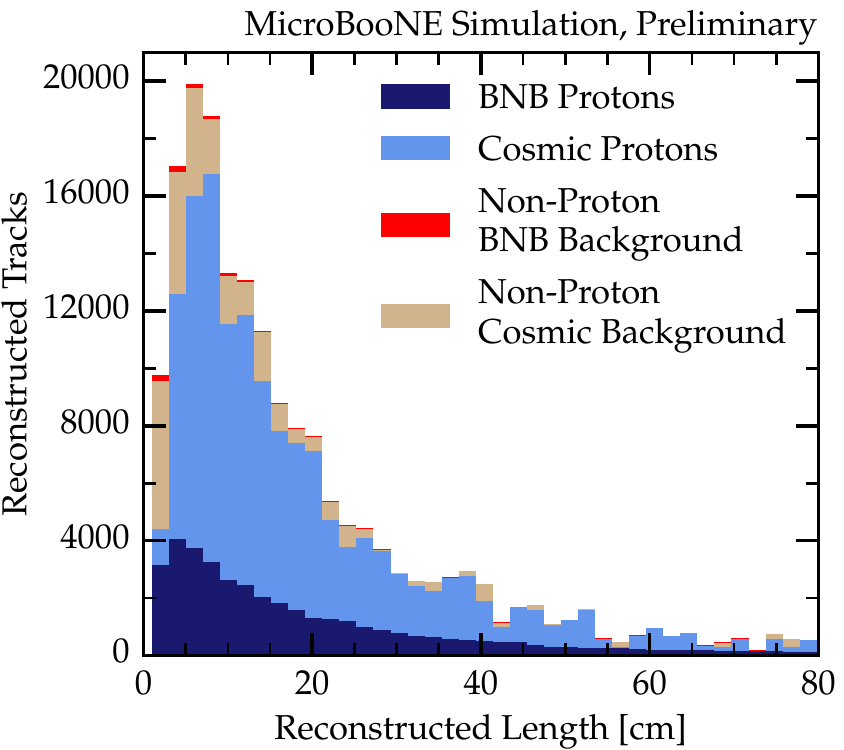}\hspace{2pc}%
  \begin{minipage}[b]{14pc}%
  \caption{Breakdown of the simulated particle types that are classified as
  protons by the boosted decision trees as a function of reconstructed track
  length. The blue filled area shows all simulated protons, both cosmic and
  neutrino-induced, and the dark blue line shows the protons from simulated BNB
  neutrino interactions. The tan filled area shows all other simulated cosmic
  tracks that are classified as protons, and the red filled area shows all
  other tracks from simulated BNB neutrino interactions that are classified as
  protons. \label{fig:bkgd}}
  \end{minipage}
\end{figure}

\begin{figure}
  \centering
  \includegraphics[scale=1.]{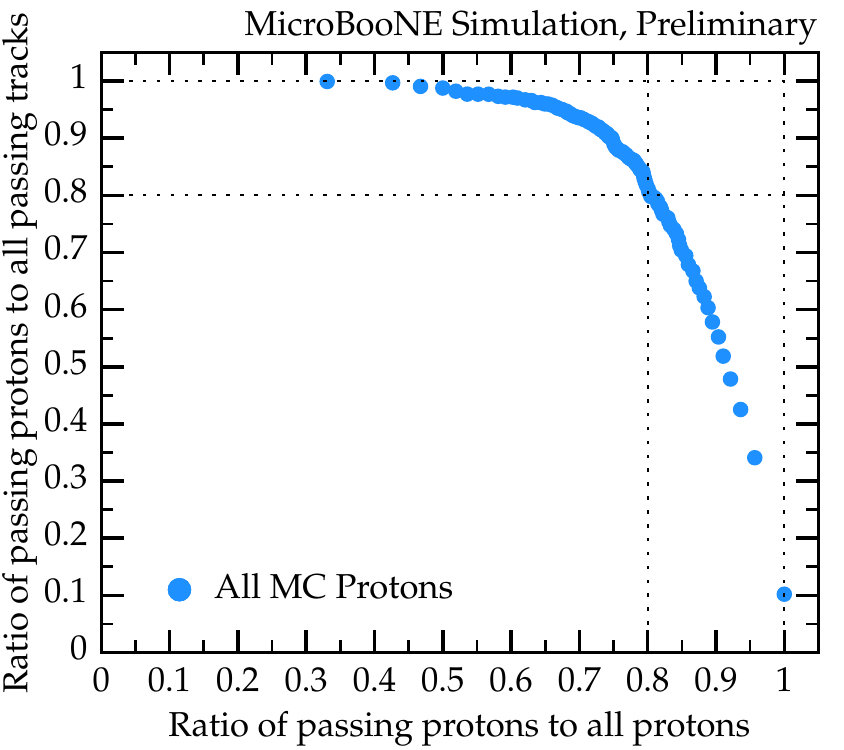}\hspace{2pc}%
  \begin{minipage}[b]{14pc}%
  \caption{The efficiency versus the purity of simulated protons selected by
  the boosted decision tree classifier for a series of proton probability cuts
  between zero and one.  \label{fig:effvpur}}
  \end{minipage}
\end{figure}

Of the reconstructed test tracks that were input to the classifier, 84\% of the
protons from simulated neutrino interactions, and 63\% of the protons from
simulated cosmic interactions were classified correctly as protons.
Figure~\ref{fig:effvke} shows the protons from simulated neutrino interactions
as a function of proton kinetic energy. Of the reconstructed test tracks that
were classified as protons, 89\% were true simulated protons (22\% neutrino
    induced protons and 67\% cosmic induced protons). Figure~\ref{fig:bkgd}
shows the breakdown of track types that are classified as protons. To maximize
efficiency or purity we can require a lower or higher proton probability from
the classifier. Figure~\ref{fig:effvpur} shows the efficiency versus purity for
different proton probability cuts in the range from zero to one.


\begin{figure}
\begin{minipage}[t]{17pc}
  \includegraphics[scale=0.2]{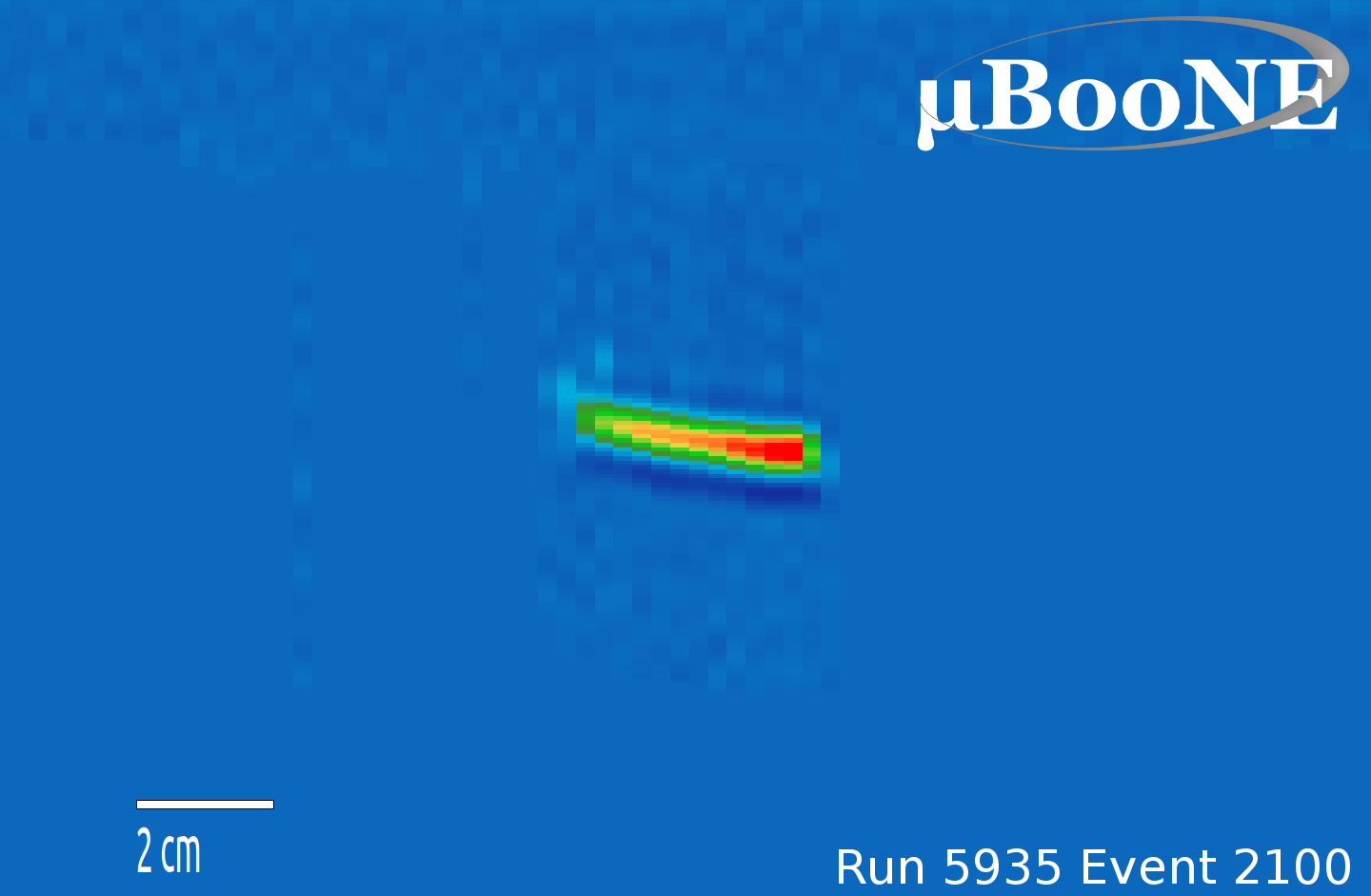}
  \caption{ Proton track candidate in MicroBooNE data. The track was selected
  by the decision tree classifier as being very likely a proton.
  \label{fig:evdnc}}
\end{minipage}\hspace{2pc}%
\begin{minipage}[t]{17pc}
  \includegraphics[scale=0.2]{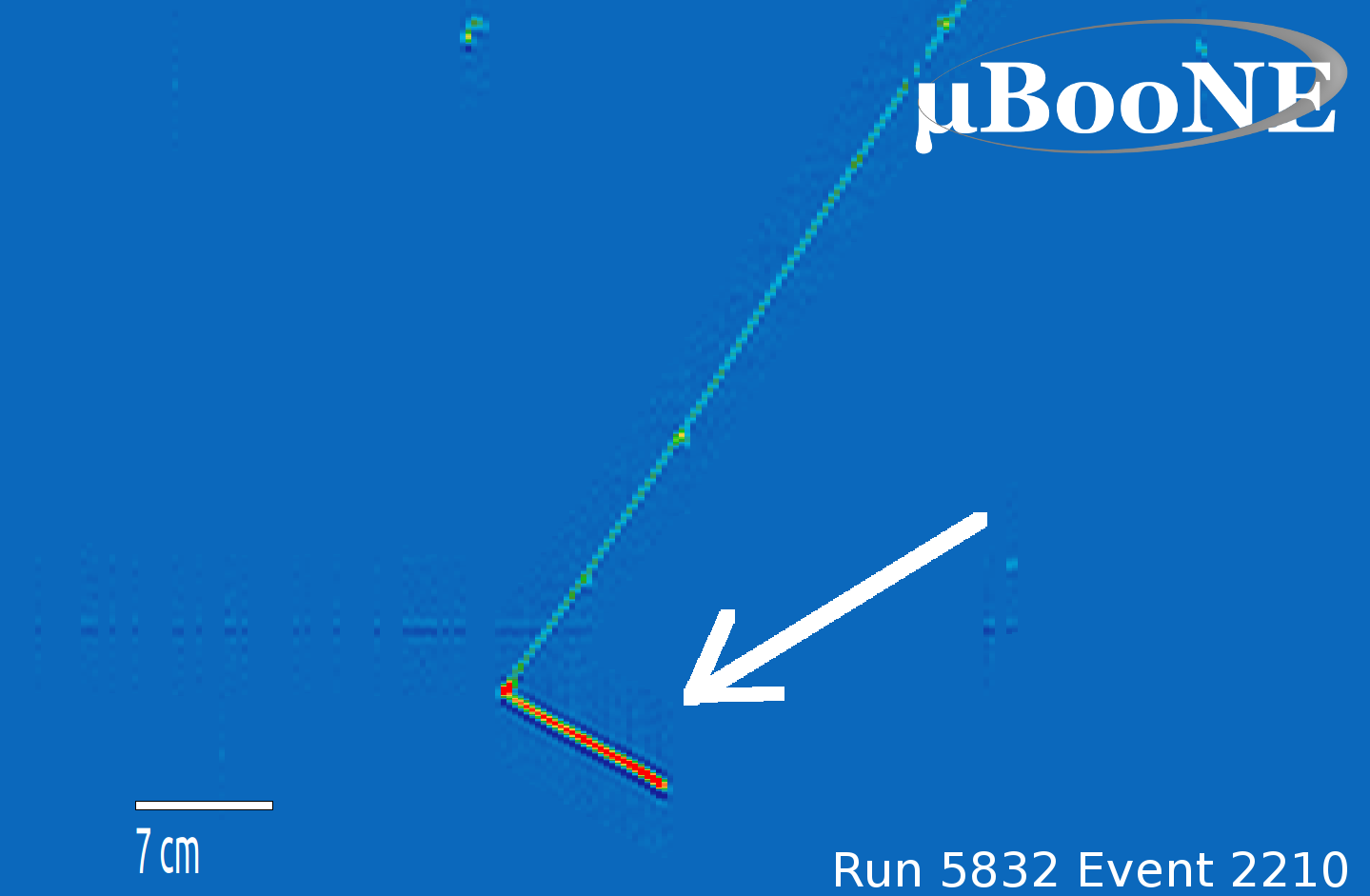}
  \caption{ Proton track candidate in MicroBooNE data. The white arrow points
  to a track that was selected by the decision tree classifier as being very
  likely a proton. \label{fig:evdcc}}
\end{minipage} 
\end{figure}

The decision tree classifier was used on a small sample of MicroBooNE data as a
performance check. Figures~\ref{fig:evdnc}~and~\ref{fig:evdcc} show tracks from
the data sample that were selected by the classifier as being very likely
protons.

\subsection{NC elastic event selection}
So far, we have kept the proton selection general to all interaction types. For
NC elastic events, we would use the output of the decision trees along with
other event information such as the total number of reconstructed tracks to select
the events of interest. This can also be used to select charged-current elastic
events with a similar efficiency to use for normalization of the NC elastic
cross section. If we are only interested in one specific topology, and do not
wish to be general, it is trivial to re-train the classifier using protons from
NC elastic interactions as the only positive input and protons from other
interactions as a background input.

\section{Conclusions \label{conclusion}}
Whether the strange quarks in the nucleon sea contribute negatively or not at
all to the spin of the nucleon is an open question. Elastic neutrino-proton
scattering offers an unique way to determine $\Delta s$ that is independent of
the assumptions required by previous measurements. The MicroBooNE liquid argon
TPC can detect low-$Q^2$ NC elastic events and is currently taking neutrino
data at Fermilab. Automated event reconstruction and selection methods are
being developed to analyze the large amount of high-resolution neutrino events
in MicroBooNE. In these proceeding, we summarize the current status of
MicroBooNE’s automatic reconstruction and identification of proton tracks in
neutral-current and charged-current interactions. The automated reconstruction
chain in MicroBooNE successfully reduces the size of the events from millions
of pixels to individual reconstructed track objects. This allows us to use
boosted decision trees for proton identification. The boosted decision tree we
presented here classifies 84\% of simulated protons from neutrino interactions
correctly and further tuning and improvement of the results is possible.

\section*{Acknowledgments}
This work was supported by the US Department of Energy, Office of Science,
Medium Energy Nuclear Physics Program.

\section*{References}
\bibliography{proceedings}

\end{document}